\documentclass[letterpaper, 10 pt, conference]{ieeeconf}  

\IEEEoverridecommandlockouts                              
\usepackage{graphicx} 
\usepackage{cite}
\usepackage{amsmath} 
\usepackage{amssymb}  
\usepackage[table,dvipsnames]{xcolor} 
\usepackage{booktabs}

\title{\LARGE \bf
Decoding Hand Kinematics from Local Field Potentials Using \\ Long Short-Term Memory (LSTM) Network
}

\author{Nur Ahmadi$^{1,2}$, \textit{Student Member, IEEE}, Timothy G. Constandinou$^{1,2}$, \textit{Senior Member, IEEE,} and \\ Christos-Savvas Bouganis$^{1}, \textit{Senior Member, IEEE}$
\thanks{*Nur Ahmadi is supported by the graduate scholarship awarded by Indonesia Endowment Fund for Education (LPDP), Republic of Indonesia.}
\thanks{$^{1}$Nur Ahmadi, Timothy G. Constandinou, and Christos-Savvas Bouganis are with the Department of Electrical and Electronic Engineering, Imperial College London, SW7 2BT, UK. Email: \{n.ahmadi16, t.constandinou, christos-savvas.bouganis\}@imperial.ac.uk}%
\thanks{$^{2}$Nur Ahmadi and Timothy G. Constandinou are additionally with the Centre for Bio-Inspired Technology, Institute of Biomedical Engineering, Imperial College London, SW7 2AZ, UK}%
\thanks{}
}

\begin{document}
\bstctlcite{IEEEexample:BSTcontrol}

\maketitle
\pubid{\begin{minipage}{\textwidth}\ \\[10pt] 
  This preprint is the ``accepted" version by IEEE NER. \copyright 2019 IEEE. Personal use of this material is permitted. Permission from IEEE must be obtained for all other uses, in any current or future media, including reprinting/republishing this material for advertising or promotional purposes, creating new collective works, for resale or redistribution to servers or lists, or reuse of any copyrighted component of this work in other works.
\end{minipage}}

\begin{abstract}
Local field potential (LFP) has gained increasing interest as an alternative input signal for brain-machine interfaces (BMIs) due to its informative features, long-term stability, and low frequency content. However, despite these interesting properties, LFP-based BMIs have been reported to yield low decoding performances compared to spike-based BMIs. In this paper, we propose a new decoder based on long short-term memory (LSTM) network which aims to improve the decoding performance of LFP-based BMIs. We compare offline decoding performance of the proposed LSTM decoder to a commonly used Kalman filter (KF) decoder on hand kinematics prediction tasks from multichannel LFPs. We also benchmark the performance of LFP-driven LSTM decoder against KF decoder driven by two types of spike signals: single-unit activity (SUA) and multi-unit activity (MUA). Our results show that LFP-driven LSTM decoder achieves significantly better decoding performance than LFP-, SUA-, and MUA-driven KF decoders. This suggests that LFPs coupled with LSTM decoder could provide high decoding performance, robust, and low power BMIs. 

\end{abstract}

\section{Introduction}
\label{Section:Introduction}
\pubidadjcol
Brain machine interfaces (BMIs) hold the promise to restore lost motor function in persons with neurological disorders (e.g. spinal cord injury) by enabling interaction with the environment through their neural activity. A key component of a BMI system is the decoder which translates neural activity (i.e. intentions) into motor commands to control external devices such as computer cursors and robotic arms, or the subject's own muscles. To date, high decoding performance BMIs have predominantly utilized spikes as the decoder's input signals which are recorded using intracortical microelectrode arrays \cite{stavisky2015high}. Despite compelling results reported in several studies \cite{shanechi2017brain,brandman2017human}, spike-based BMIs face two major challenges towards their clinically viable translation \cite{jackson2017decoding}. First, the number of recorded spike signals declines over time, which can degrade the performance of BMI decoder. This long-term instability is thought to be caused by scar tissue formation around the electrodes and micromotion of the electrodes \cite{jackson2017decoding,hwang2013utility}. Second, due to high sampling rate ($>$10\,kHz), detecting and sorting spikes in an implanted circuitry or transmitting the raw data require high power consumption, which will in turn increase heat and limit the device's battery lifetime \cite{jackson2017decoding}.  

One approach to address these challenges is to utilize another signal modality within neural activity, namely local field potential (LFP), as an alternative input signal for BMI decoder. LFP is a low frequency extracellular voltage thought to be mainly generated from postsynaptic currents and reflect activity of population of neurons in the vicinity of the electrodes. LFP is thus believed to be less sensitive to scar formation and micromotion of the electrodes \cite{jackson2017decoding}. Several studies have shown that LFPs are more stable than spikes and that considerable amount of movement-related information contained within LFPs are still present even after spikes are lost \cite{flint2013long,wang2014long,stavisky2015high}. In addition, LFPs can be processed at significantly lower sampling rate than spikes, which translates into lower power consumption and lower complexity (i.e. memory requirements). However, despite these interesting properties, LFP-based BMIs have been shown to yield low decoding performances compared to spike-based BMIs \cite{bansal2011decoding,flint2012accurate,flint2013long,hwang2013utility,stavisky2015high}. Most of LFP-based BMIs employ Kalman filter (KF) decoder which assumes linear and Gaussian distribution on both the observation and state dynamics. However, since LFPs exhibit nonlinear, non-stationary, and non-Gaussian characteristics \cite{menzer2010characterization}, aforementioned assumptions could lead to suboptimal results. Therefore, a more flexible and effective decoding method is required to improve the decoding performance of LFP-based BMIs.

One promising solution to this issue is recurrent neural networks with long short-term memory (LSTM) architecture, which do not require any assumption on the data. In fact, LSTMs have become the state-of-the-art methods for various applications ranging from speech recognition and synthesis, language modeling and translation, to audio and video analysis \cite{greff2017lstm}. Despite the successes in a variety of fields, to our knowledge, however, LSTMs have not been applied to LFP-based BMI decoding.  

\pubidadjcol 
\begin{figure*}[!tp]
\centering
\includegraphics[trim=0.1in 0in 0.1in 0in, clip=true,width=0.9\textwidth]{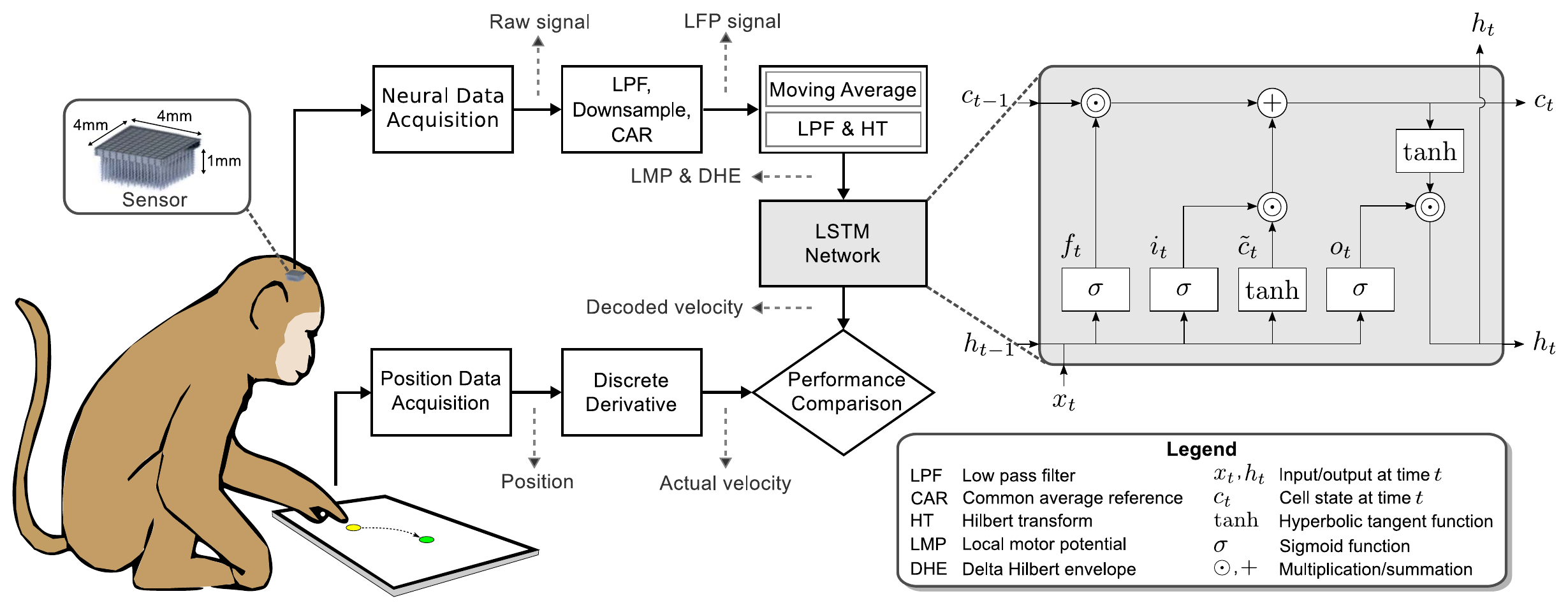}
\caption{Block diagram of LFP-based BMI decoding using an LSTM network. The larger gray-shaded block illustrates a detailed schematic of an LSTM.}
\label{fig:lfp_bmi_blockdiagram}
\end{figure*}

In light of this, in the present paper, we propose an LSTM-based decoder for decoding hand kinematics from multichannel LFPs recorded intracortically from non-human primate (NHP) during free-reaching tasks. We then compare offline decoding performance of the proposed LSTM decoder to a commonly used Kalman filter (KF) decoder. Additionally, we also benchmark the decoding performance between LFP-driven LSTM decoder and KF decoder driven by two types of spike signals: single-unit activity (SUA) and multi-unit activity (MUA).

\section{Methods}
\label{Section:Methods}
\subsection{Neural Recording and Behavioral Task}
In this study, we used two datasets (dataset I: I\_20170124\_01; dataset II: I\_20170127\_03) from a public neural data recorded from primary motor cortex (M1) area of an adult male Rhesus macaque monkey (\textit{Macaca mulata}) by Sabes Lab \cite{odoherty2017nonhuman}. Recording duration for dataset I and II were 9 and 12 minutes, respectively. The neural data was recorded using 96-channel silicon microelectrode array (Utah array with 400$\,\mu$m inter-electrode distance and 1$\,$mm electrode length) while the monkey was performing self-paced reaching tasks without inter-trial intervals within 8-by-8 square grid. The recording was referenced to a silver wire placed under the dura (several cm away from the electrodes). The neural data was sampled at 24.4$\,$kHz and filtered with a 4$^{th}$ oder low-pass filter at 7.5$\,$kHz. Fingertip position in $x$-$y$ coordinates was sampled at 250$\,$Hz. Velocity was then computed from position by using discrete derivative. A more detailed description of the experiment setup is given in \cite{makin2017superior}. 

\subsection{Neural Signal Processing}
 LFPs were obtained by low pass filtering raw neural data using 4$^{th}$ order Butterworth filter at 300$\,$Hz and then downsampling them to 1$\,$kHz. LFPs have been shown to contain common noises, partly arising from the use of a single, distal reference (unipolar) \cite{xinyu2017adaptive}. To remove these common noises, we performed common average reference (CAR) and subtracted it from LFP signal in each channel. At each time instant, CAR was computed by averaging LFP signals across all channels. Two LFP features, local motor potential (LMP) \cite{flint2012accurate,flint2013long,stavisky2015high} and delta Hilbert envelope (DHE), were extracted. LMPs were computed from LFPs by using time-domain moving average filter. DHEs were calculated firstly by low pass filtering LFPs with 4$^{th}$ order Butterworth filter at 4$\,$Hz and then computing the envelope (i.e. instantaneous amplitude) of their analytic signals through Hilbert transform (HT). We computed Hilbert envelope only from delta band (0-4$\,$Hz) since other frequency bands in our data did not contain significant information as observed through time-frequency decomposition. Both LMP and DHE features were computed within overlapping windows that slide every 4$\,$ms to match the timescale of hand kinematics. We used window widths ranging from 32$\,$ms to 512$\,$ms with an increment of 32$\,$ms. For performance evaluation, we selected a window width and LFP features that yielded the best decoding performance on validation set. The processing steps of our LFP-based BMI decoding is illustrated in Fig.~\ref{fig:lfp_bmi_blockdiagram}.

Spikes were extracted from each channel by band pass filtering (from 500 to 5000\,Hz) raw neural data and then detecting the filtered signal amplitudes that crossed a predetermined threshold value. All the detected spikes within single channel are referred to as multi-unit activity (MUA). To extract single-unit activity (SUA), the detected spikes were sorted (i.e. classified) into distinct putative single units. More detailed information on spike detection and sorting processes can be found in \cite{makin2017superior}. In this study, units with spike rates below 0.5\,Hz were excluded \cite{makin2017superior} thus leaving only 130 (124) SUAs and 91 (94) MUAs for dataset I (II). We computed spike counts within different overlapping bin/window widths (same range as of LFPs) sliding every 4\,ms (corresponding to the kinematics time scale). Overlapping bin was used because it has been demonstrated to yield better decoding performance than non-overlap bin \cite{koyama2010comparison,ahmadi2018spike}. Similar to that of LFP-based decoding, we selected a bin width that resulted in the best decoding performance on validation set.

\subsection{Long Short-Term Memory (LSTM) Network} 
Long short-term memory (LSTM) is a type of recurrent neural networks (RNNs) that was developed by Hochreiter and Schmidhuber in 1997 \cite{hochreiter1997long}. It has successfully addressed the vanishing or exploding gradient problem encountered when training traditional RNNs. LSTM is very well suited to many sequential data problems as it is capable of learning long-term temporal dependencies through gating mechanism. A commonly used variant of LSTM units contains three gates (forget, input, and output) and a memory cell that control the flow of information. In this study, we used this LSTM variant where its components' state at time instant $t$ can be described by:
\begin{equation}
    \label{eqn:matrix1}
    \begin{array}{r@{}l}
    f_t &{}= \sigma (W_fx_t + U_fh_{t-1} + b_f) \\
    i_t &{}= \sigma (W_ix_t + U_ih_{t-1} + b_i) \\
    \tilde{c}_t &{}= \tanh (W_cx_t + U_ch_{t-1} + b_c) \\
    c_t &{}= f_t \odot c_{t-1} + i_t \odot \tilde{c}_t \\
    o_t &{}= \sigma (W_ox_t + U_oh_{t-1} + b_o) \\
    h_t &{}= \tanh (c_t) \odot o_t 
    \end{array}
\end{equation}
where $x,h,f,i,o,c$ represent the input, output, forget gate, input gate, output gate, and memory cell, respectively; $\sigma$ and $\tanh$ denote sigmoid and hyperbolic tangent activation functions, respectively; $\odot$ denotes element-wise multiplication. $W, U, b$ are weight matrices and bias vector parameters that need to be learned during training.  

Training an LSTM decoder requires setting up configuration parameters called hyperparameters, which in this study are given in Table~\ref{tab:hyperparameter}. The hyperparameter configuration was tuned through a Bayesian optimization library (Hyperopt) \cite{bergstra2015hyperopt} with 200 iterations. The LSTM decoder was implemented using Keras framework with TensorFlow backend and trained using RMSprop optimizer and root mean squared error (RMSE) loss function.
\begin{table}[!tp]
\centering
\caption{Hyperparameters for LSTM decoder}
\footnotesize
\begin{tabular}{l c }
\toprule
Hyperparameter & Values\\
\midrule
Number of units   & $\{50, 75, 100, \cdots , 200\}$  \\ 
Number of epochs   & $\{2, 3, 4, \cdots , 8\}$  \\ 
Batch size   & $\{32, 64, 96, 128\}$  \\ 
Dropout rate     & $\{0, 0.1, 0.2, 0.3, 0.4\}$  \\
Learning rate   & $\{0.001, 0.0015, 0.002, \cdots , 0.003\}$  \\
\bottomrule
\end{tabular}
\label{tab:hyperparameter}
\end{table}

\subsection{Performance Evaluation and Metrics}
To evaluate the performance of the LSTM decoder, we used two common metrics, namely, root mean squared error (RMSE) and Pearson's correlation coefficient. These metrics were computed between decoded and actual velocities on $x$- and $y$- coordinates. We split each dataset into $k=10$ subdatasets which were then categorized into a training set (concatenation of $k-2$ subdatasets), validation set (1 subdataset), and testing set (1 subdataset). The training set was used to find values of the LSTM parameters ($W, U, b$). The validation set (from dataset I) was used to find the optimal configuration of LSTM hyperparameters, LFP features, window widths, and lags between LFP/spike features and the kinematics that minimized average RMSE across $x$-$y$ coordinates. This configuration was then used for evaluating the performance of LSTM decoder on testing sets. We iterated the performance evaluation on $k$ different testing sets for each dataset.

\section{Results}\label{Section:Results}
According to our evaluation on validation set (dataset I), the following configuration of LSTM hyperparameters led to smallest average RMSE: number of layer = 1, number of timesteps = 2, number of units = 100, number of epochs = 6, batch size = 32, dropout rate = 0.2, and learning rate = 0.001. This configuration was then used to find the optimal LFP features and window width using the validation set (dataset I) and to benchmark the performance of LSTM decoder against KF decoder using the testing sets (dataset I and II).

\subsection{LFP Feature and Window Width Selection}
\begin{figure}[!tp]
\centering
\includegraphics[trim=0.1in 0in 0.1in 0in, clip=true,width=0.825\columnwidth]{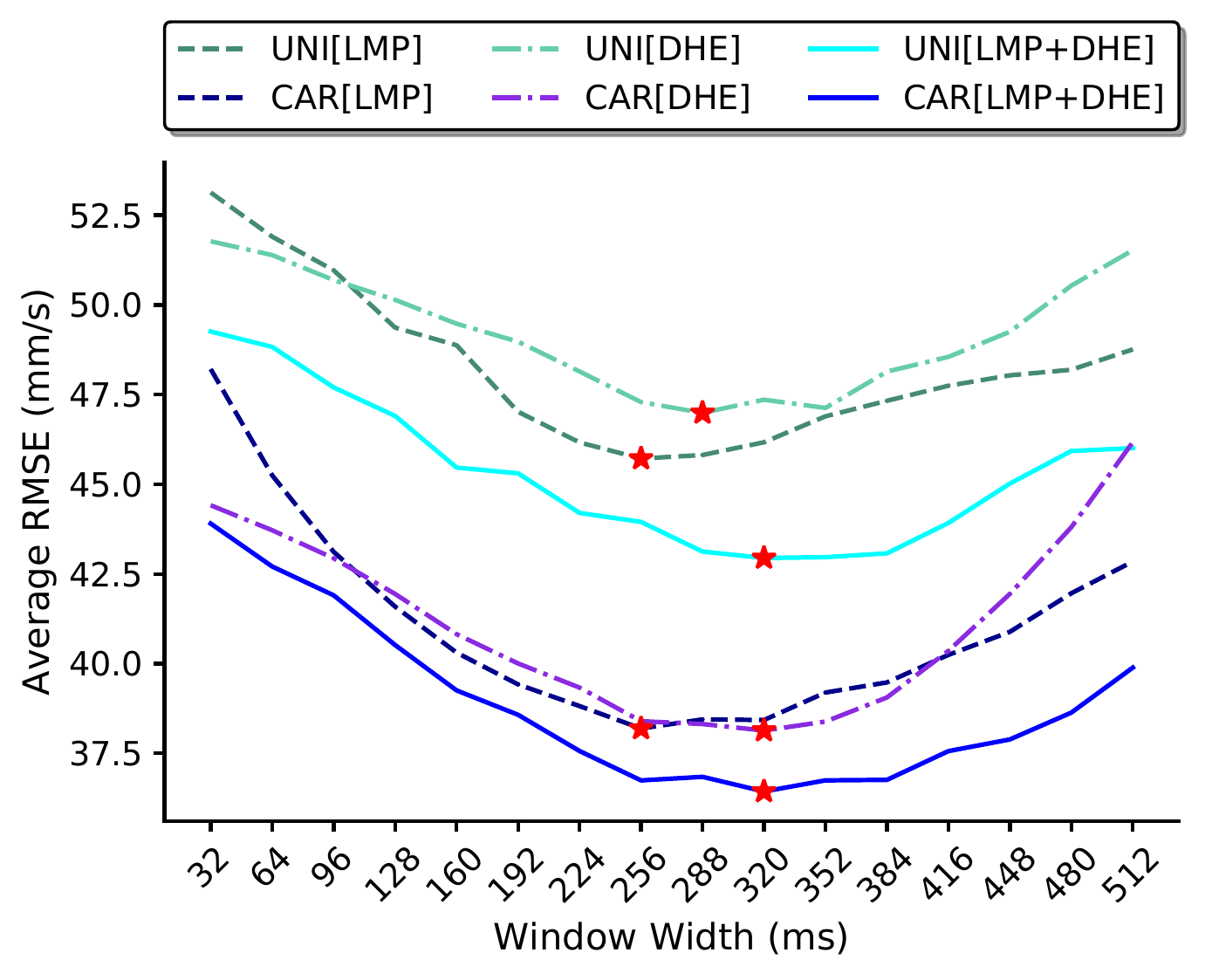}
\caption{Comparison of decoding performance of LSTM decoder under different LFP features and window widths. Red stars indicate window widths that yielded smallest average RMSEs on validation set of dataset I.}
\label{fig:lfp_decode_window}
\end{figure}

The LFP features used for comparison were LMP, DHE, and combination of both, which were extracted from two reference schemes (unipolar and CAR). According to our results, CAR-based LFP features led to superior decoding performance than unipolar-based LFP features. LMP, on average, resulted in better performance than DHE. Combination of both LMP and DHE features yielded improved performance than that of single LFP feature. In regards to different window widths ($\{32,64,96,\cdots,512\}\,$ms), we observed a similar trend across different LFP features. The decoding performance increased (smaller average RMSE) as the window width increased up to a particular point where above this point the performance would decrease. This particular point corresponds to a window width within a range of $256-320\,$ms. The summary of LSTM decoder performance with respect to different LFP features and window widths is depicted in Fig.~\ref{fig:lfp_decode_window}. For performance benchmark against KF decoder on testing sets, we selected a combination of LMP and DHE features extracted from CAR-based LFPs with $320\,$ms window width. Through the same procedure, we selected a combination of LMP and DHE features extracted from unipolar-based LFPs with $256\,$ms for KF decoder (figure is not shown due to page limitation).

\subsection{Window Width Selection for Spike-driven Decoders}
\begin{figure}[!tp]
\centering
\includegraphics[trim=0.1in 0in 0.1in 0in, clip=true,width=0.825\columnwidth]{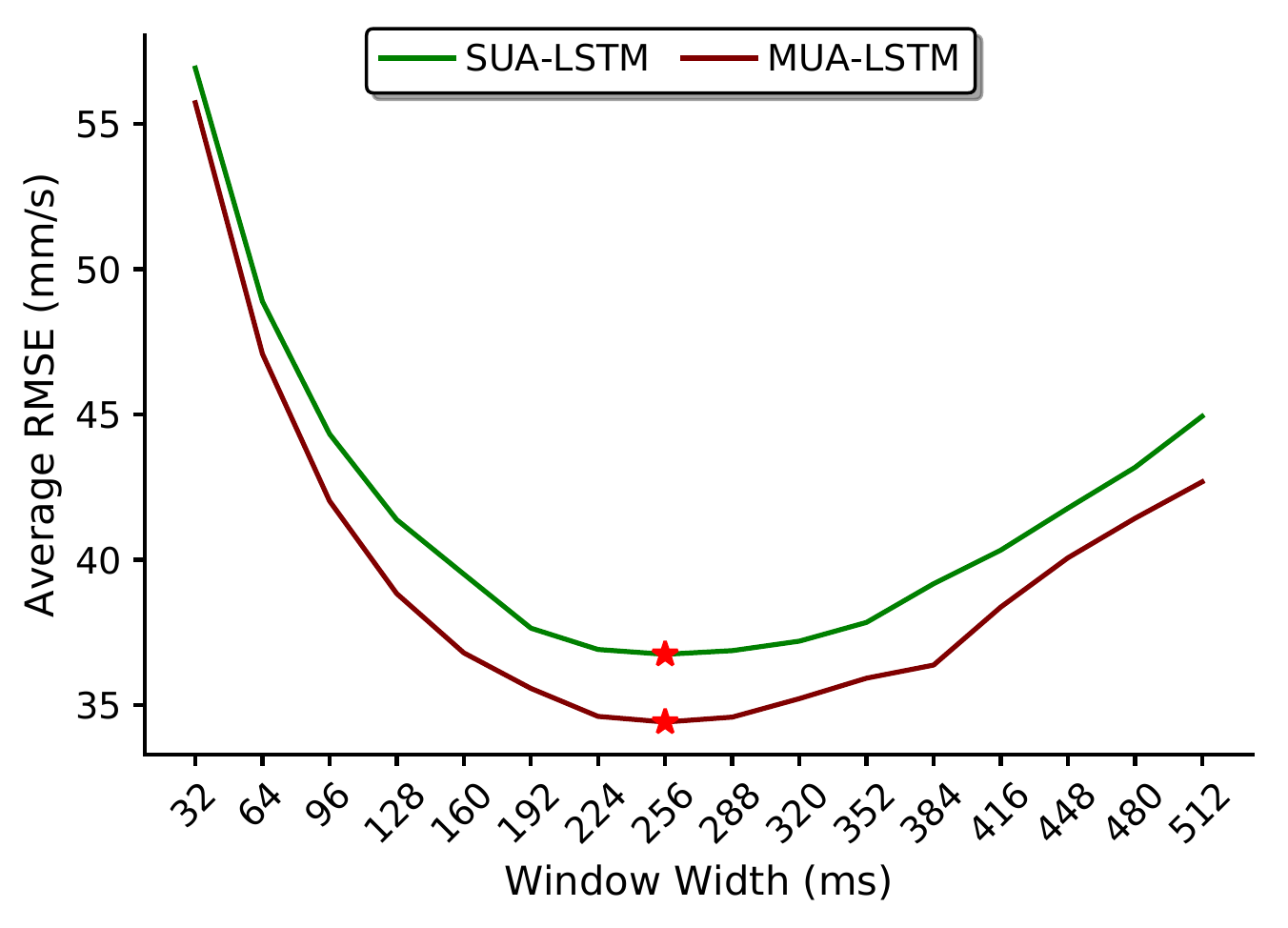}
\caption{Comparison of decoding performance of LSTM decoder driven by SUA and MUA under different window widths. Red stars indicate window widths that yielded smallest average RMSEs on validation set of dataset I.}
\label{fig:spike_decode_window}
\end{figure}

We also investigated the impact of different window widths to decoding performance of LSTM and KF decoders driven by spike signals (SUA and MUA). A similar trend as in LFP-driven decoders was also observed in the cases of spike-driven LSTM and KF decoders. The decoding performance improved as the window width increased and at a particular point the performance would decline. The summary of spike-driven LSTM decoder performance with respect to different window widths is shown in Fig.~\ref{fig:spike_decode_window}. Due to page limitation, the summary of spike-driven KF decoder is not shown. For performance comparison on testing sets, we selected a window width of $256\,$ms for both SUA- and MUA-driven LSTM decoders. Window widths of $256\,$ms and $288\,$ms were selected for SUA- and MUA-driven KF decoders, respectively.

\subsection{Decoding Performance Comparison}
\begin{figure}[!tp]
\centering
\includegraphics[trim=0.1in 0in 0.1in 0in, clip=true,width=\columnwidth]{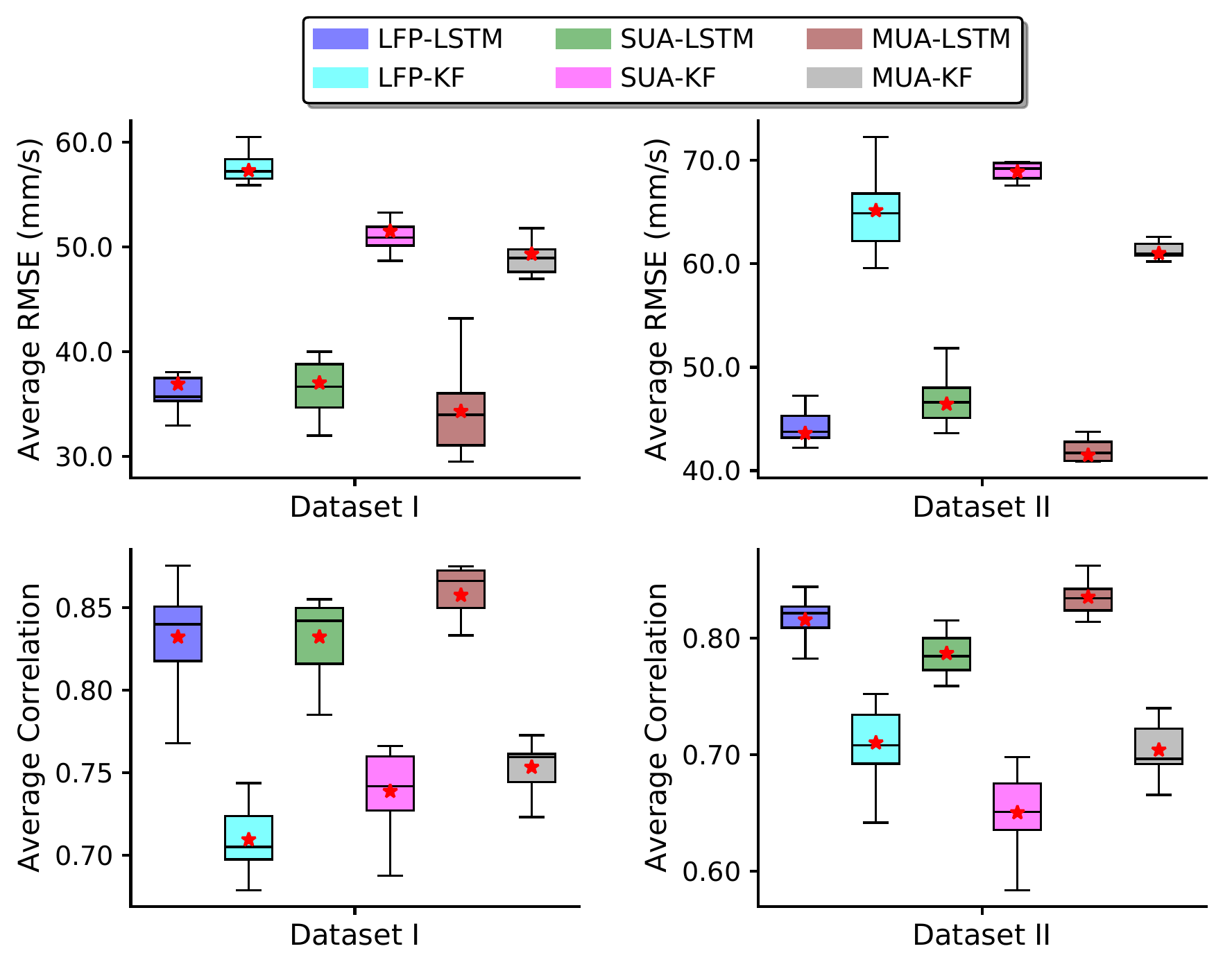}
\caption{Decoding performance comparison of LSTM and KF decoders under different input signals. Black horizontal lines inside the boxes represent the medians; red stars represent the means; colored boxes denote the interquartile ranges; whiskers extend $1.5\times$ from upper and lower quartiles.}
\label{fig:boxplot_decode_all}
\end{figure}

\begin{figure}[!tp]
\centering
\includegraphics[trim=0.1in 0in 0.1in 0in, clip=true,width=\columnwidth]{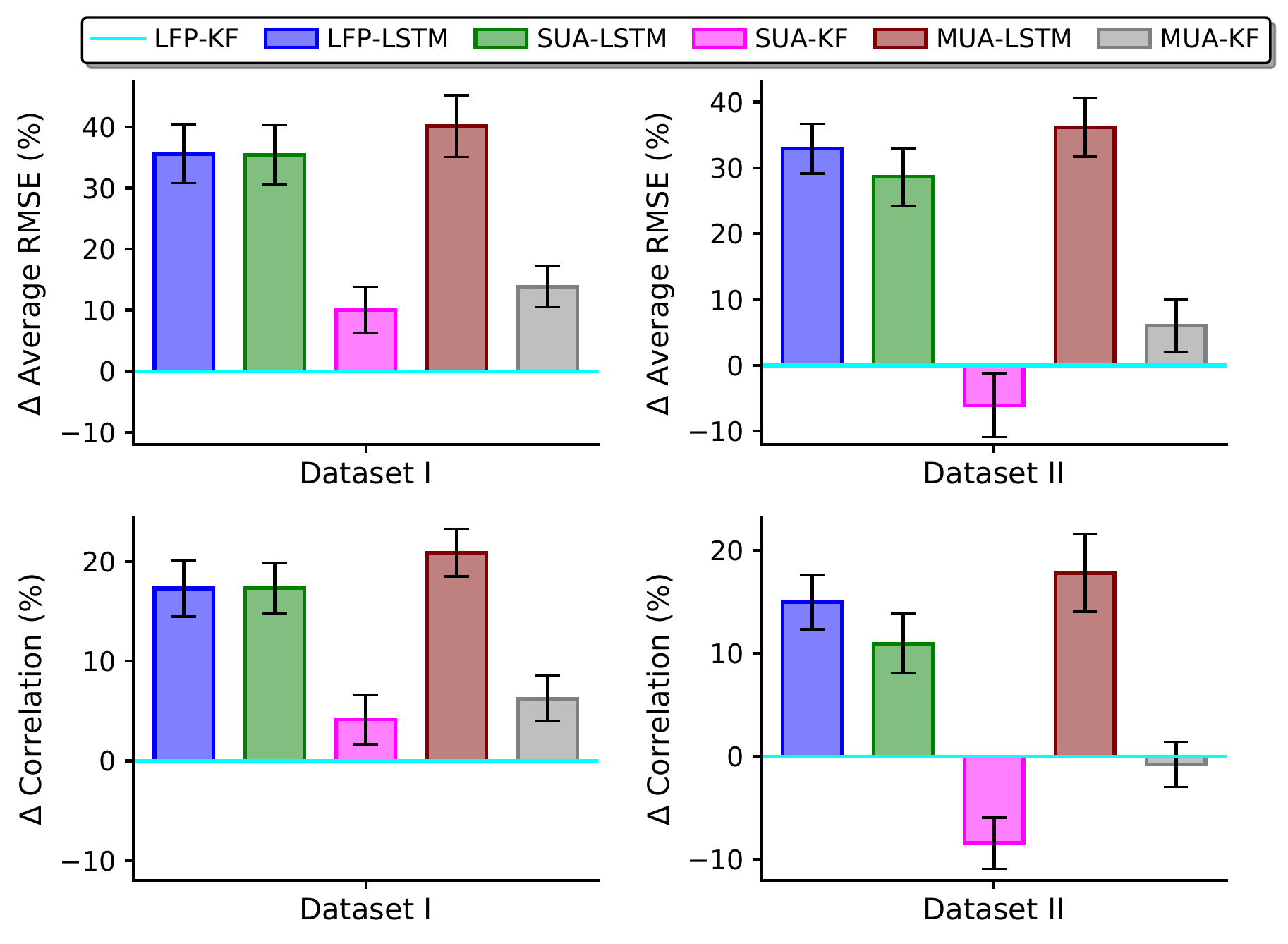}
\caption{Performance comparison of different decoders relative to LFP-KF. A Positive (negative) value indicates performance improvement (decrease). Black error bars represent 95\% confidence intervals.}
\label{fig:lfp_lstm_decode_improve}
\end{figure}

Firstly, we compared the decoding performance of our proposed LFP-driven LSTM decoder to LFP-, SUA-, and MUA-driven KF decoder (referred to as LFP-LSTM, LFP-KF, SUA-KF, and MUA-KF respectively). As shown in Fig.~\ref{fig:boxplot_decode_all}, LFP-LSTM achieved significantly better decoding performance than LFP-KF, SUA-KF, and MUA-KF measured in both metrics (RMSE and correlation) for both datasets (I and II). LFP-LSTM yielded average RMSE of $36.90 \pm 2.85$ $(43.60 \pm 2.31)$ mm/s and average correlation of $0.83 \pm 0.02$ $(0.82 \pm 0.02)$ for dataset I (II). This average RMSE value corresponded to performance improvement of $35.59 \pm 4.77$ $(32.93 \pm 3.78)\%$ for dataset I (II) with respect to LFP-KF as shown in Fig.~\ref{fig:lfp_lstm_decode_improve}. In term of average correlation, the performance improvement made by LFP-LSTM over LFP-KF was $17.32 \pm 2.85$ $(14.98 \pm 2.65)\%$ for dataset I (II).

\begin{figure}[!tp]
\centering
\includegraphics[trim=0.1in 0in 0.1in 0in, clip=true,width=0.75\columnwidth]{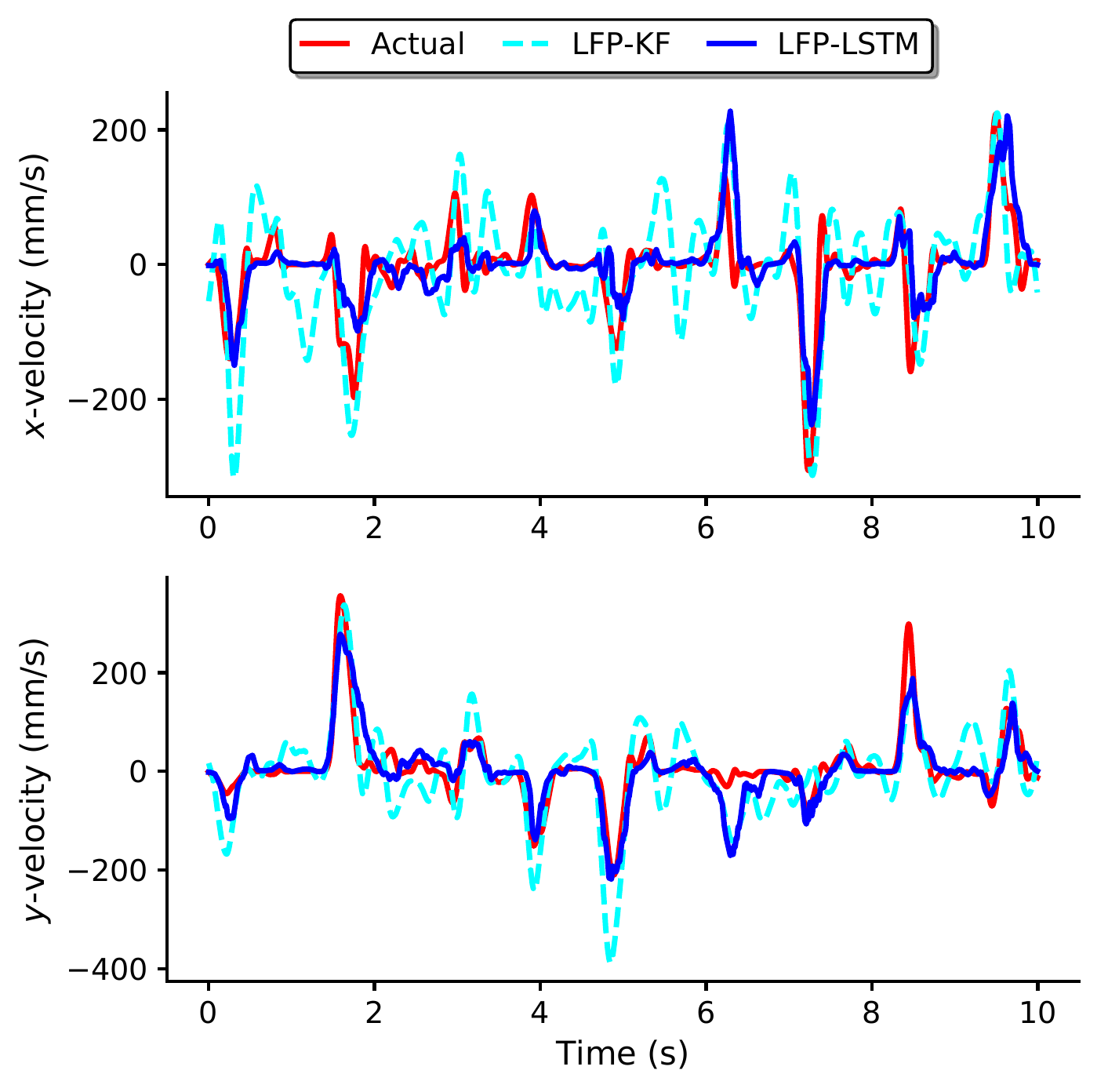}
\caption{Examples of actual and decoded velocities in $x$- and $y$-directions from dataset II.}
\label{fig:compare_decode_signal}
\end{figure}

In addition, we also compared the decoding performance of LFP-LSTM to SUA- and MUA-driven LSTM decoder (referred to as SUA-LSTM and MUA-LSTM). The results show that LFP-LSTM outperformed SUA-LSTM but was outperformed by MUA-LSTM for both datasets. The overall performance comparison among different decoding methods is shown in Fig.~\ref{fig:boxplot_decode_all}. We plotted performance improvement/decrease of different decoders with respect to LFP-KF as can be seen in Fig.~\ref{fig:lfp_lstm_decode_improve}. MUA-LSTM yielded largest performance improvement of $40.15 \pm 5.07$ $(36.15 \pm 4.42)\%$ in average RMSE and $20.91 \pm 2.40$ $(17.81 \pm 3.78)\%$ in average correlation for dataset I (II). SUA-KF showed improved performance than LFP-KF in dataset I but worse performance in dataset II. Examples of actual and decoded velocities (in $x$ and $y$ directions) from LFP-LSTM and LFP-KF are given in Fig.~\ref{fig:compare_decode_signal}. For the sake of readability, only decoded velocities from these two decoders are shown. 

\section{Discussion}
\label{Section:Discussion}
In this study, we propose a new decoder based on LSTM network to predict (offline) hand kinematics from multichannel LFPs of an NHP subject. We have shown that LFP-driven LSTM decoder significantly outperforms KF decoder driven by LFP, SUA, and MUA. This demonstrates the effectiveness of LSTM decoder in capturing nonlinear and complex relationship between LFPs and hand kinematics. On the other hand, KF decoder with its inherent assumptions results in poor performance, especially for LFPs that exhibit high spatial correlation \cite{xinyu2017adaptive}. LFPs recorded from unipolar reference are often contaminated by spatially correlated noises, which affects the decoding performance. These noises can be eliminated by using common average reference (CAR) employed prior to LFP feature extraction process. Our results show that CAR can lead to improved decoding performance. Our results also show that LMP contains significant amount of movement-related information as indicated by good decoding performance compared to other LFP features, which is in good agreement with \cite{stavisky2015high}. However, our results differs from \cite{stavisky2015high} in that delta band of LFP is more informative than the higher bands. By combining LMP with DHE as the input for LSTM decoder, the performance of LSTM decoder can be improved. Prior studies demonstrated that LFP-driven decoder performed slightly worse or comparable to spike-driven decoder \cite{stavisky2015high,hwang2013utility}. Using hybrid LFP-spike signals, they could achieve better decoding performance than that of spikes only. Here, we show that using only LFP signals, our proposed LFP-driven LSTM decoder can significantly outperform spike-driven KF decoder.

Performance comparison using the same LSTM decoder with different input signals (LFP, SUA, and MUA) show that MUA-LSTM achieves the best decoding performance, followed by LFP-LSTM and SUA-LSTM, respectively. This may indicate that MUA contains more movement-related information than that of SUA. Better performance of MUA over SUA for both KF and LSTM decoders obtained from this study offers a new perspective to the debate whether or not spikes should be sorted \cite{todorova2014sort,christie2014comparison}. However, the changes of spike amplitude and waveform over time pose critical challenges in adjusting the threshold crossing value for spike detection, sorting and tracking the spikes into the same putative single units. Furthermore, the high sampling rate required for these spike processing demands high power consumption that hinders the implementation of wireless, scalable, and implantable BMIs. LFP offers attractive properties to address these challenges: long-term stability and low-frequency content (i.e. low sampling rate). In this study, we only use two datasets corresponding two recording sessions (3 days gap between the sessions). For future work, we will investigate the stability and robustness of LFP-, SUA-, and MUA-driven LSTM decoders over long period of time.

Overall, our results suggest that LFP indeed contains a rich movement-related information, which corroborates the idea that LFP is a promising alternative signal input for BMIs. Along with their stable and low-frequency properties, LFPs coupled with LSTM decoder could potentially provide high decoding performance, robust, and low power BMIs.

\section*{Acknowledgement}
We thank J. E. O'Doherty and P. N. Sabes for making their data publicly available.

\bibliographystyle{IEEEtran}
\bibliography{IEEEabrv,myreferences}

\begin{thebibliography}{10}
\providecommand{\url}[1]{#1}
\csname url@samestyle\endcsname
\providecommand{\newblock}{\relax}
\providecommand{\bibinfo}[2]{#2}
\providecommand{\BIBentrySTDinterwordspacing}{\spaceskip=0pt\relax}
\providecommand{\BIBentryALTinterwordstretchfactor}{4}
\providecommand{\BIBentryALTinterwordspacing}{\spaceskip=\fontdimen2\font plus
\BIBentryALTinterwordstretchfactor\fontdimen3\font minus
  \fontdimen4\font\relax}
\providecommand{\BIBforeignlanguage}[2]{{%
\expandafter\ifx\csname l@#1\endcsname\relax
\typeout{** WARNING: IEEEtran.bst: No hyphenation pattern has been}%
\typeout{** loaded for the language `#1'. Using the pattern for}%
\typeout{** the default language instead.}%
\else
\language=\csname l@#1\endcsname
\fi
#2}}
\providecommand{\BIBdecl}{\relax}
\BIBdecl
\renewcommand{\BIBentryALTinterwordstretchfactor}{4}

\bibitem{stavisky2015high}
S.~D. Stavisky \emph{et~al.}, ``A high performing brain--machine interface
  driven by low-frequency local field potentials alone and together with
  spikes,'' \emph{J. Neural Eng.}, vol.~12, no.~3, p. 036009, 2015.

\bibitem{shanechi2017brain}
M.~M. Shanechi, ``Brain--machine interface control algorithms,'' \emph{IEEE
  Trans. Neural Syst. Rehabil. Eng}, vol.~25, no.~10, pp. 1725--1734, 2017.

\bibitem{brandman2017human}
D.~M. Brandman \emph{et~al.}, ``Human intracortical recording and neural
  decoding for brain-computer interfaces,'' \emph{IEEE Trans. Neural Syst.
  Rehabil. Eng}, vol.~25, no.~10, pp. 1687--1696, 2017.

\bibitem{jackson2017decoding}
A.~Jackson and T.~M. Hall, ``Decoding local field potentials for neural
  interfaces,'' \emph{IEEE Trans. Neural Syst. Rehabil. Eng}, vol.~25, no.~10,
  pp. 1705--1714, 2017.

\bibitem{hwang2013utility}
E.~J. Hwang and R.~A. Andersen, ``The utility of multichannel local field
  potentials for brain--machine interfaces,'' \emph{J. Neural Eng.}, vol.~10,
  no.~4, p. 046005, 2013.

\bibitem{flint2013long}
R.~D. Flint \emph{et~al.}, ``Long term, stable brain machine interface
  performance using local field potentials and multiunit spikes,'' \emph{J.
  Neural Eng.}, vol.~10, no.~5, p. 056005, 2013.

\bibitem{wang2014long}
D.~Wang \emph{et~al.}, ``Long-term decoding stability of local field potentials
  from silicon arrays in primate motor cortex during a {2D} center out task,''
  \emph{J. Neural Eng.}, vol.~11, no.~3, p. 036009, 2014.

\bibitem{bansal2011decoding}
A.~K. Bansal \emph{et~al.}, ``Decoding {3D} reach and grasp from hybrid signals
  in motor and premotor cortices: spikes, multiunit activity, and local field
  potentials,'' \emph{J. Neurophysiol.}, vol. 107, no.~5, pp. 1337--1355, 2011.

\bibitem{flint2012accurate}
R.~D. Flint \emph{et~al.}, ``Accurate decoding of reaching movements from field
  potentials in the absence of spikes,'' \emph{J. Neural Eng.}, vol.~9, no.~4,
  p. 046006, 2012.

\bibitem{menzer2010characterization}
D.~L. Menzer \emph{et~al.}, ``Characterization of trial-to-trial fluctuations
  in local field potentials recorded in cerebral cortex of awake behaving
  macaque,'' \emph{J. Neurosci. Methods}, vol. 186, no.~2, pp. 250--261, 2010.

\bibitem{greff2017lstm}
K.~Greff \emph{et~al.}, ``{LSTM}: A search space odyssey,'' \emph{IEEE Trans.
  Neural Syst. Rehabil. Eng}, vol.~28, no.~10, pp. 2222--2232, 2017.

\bibitem{odoherty2017nonhuman}
J.~E. O'doherty \emph{et~al.}, ``Nonhuman primate reaching with multichannel
  sensorimotor cortex electrophysiology,'' 2017, doi:10.5281/zenodo.583331.

\bibitem{makin2017superior}
J.~G. Makin \emph{et~al.}, ``Superior arm-movement decoding from cortex with a
  new, unsupervised-learning algorithm,'' \emph{J. Neural Eng.}, vol.~15,
  no.~2, p. 026010, 2018.

\bibitem{xinyu2017adaptive}
L.~Xinyu \emph{et~al.}, ``Adaptive common average reference for in vivo
  multichannel local field potentials,'' \emph{Biomed. Eng. Letters}, vol.~7,
  no.~1, pp. 7--15, 2017.

\bibitem{koyama2010comparison}
S.~Koyama \emph{et~al.}, ``Comparison of brain--computer interface decoding
  algorithms in open-loop and closed-loop control,'' \emph{J. Comput.
  Neurosci.}, vol.~29, no. 1-2, pp. 73--87, 2010.

\bibitem{ahmadi2018spike}
N.~Ahmadi \emph{et~al.}, ``Spike rate estimation using bayesian adaptive kernel
  smoother ({BAKS}) and its application to brain machine interfaces,'' in
  \emph{Proc. of the Ann. Int. Conf. of the IEEE Engineering in Medicine and
  Biology Society}.\hskip 1em plus 0.5em minus 0.4em\relax IEEE, 2018, pp.
  2547--2550.

\bibitem{hochreiter1997long}
S.~Hochreiter and J.~Schmidhuber, ``Long short-term memory,'' \emph{Neural
  Computation}, vol.~9, no.~8, pp. 1735--1780, 1997.

\bibitem{bergstra2015hyperopt}
J.~Bergstra \emph{et~al.}, ``Hyperopt: a python library for model selection and
  hyperparameter optimization,'' \emph{Comput. Sci. Discovery}, vol.~8, no.~1,
  p. 014008, 2015.

\bibitem{todorova2014sort}
S.~Todorova \emph{et~al.}, ``To sort or not to sort: the impact of
  spike-sorting on neural decoding performance,'' \emph{J. Neural Eng.},
  vol.~11, no.~5, p. 056005, 2014.

\bibitem{christie2014comparison}
B.~P. Christie \emph{et~al.}, ``Comparison of spike sorting and thresholding of
  voltage waveforms for intracortical brain--machine interface performance,''
  \emph{J. Neural Eng.}, vol.~12, no.~1, p. 016009, 2014.

\end{thebibliography}

\end{document}